\documentclass[11pt]{article}

\usepackage{amsmath,amssymb}

\makeatletter

 \@addtoreset{equation}{section}
\makeatother

\newcommand{\del}{\partial}

\newcommand{\half}{\frac{1}{2}}

\newcommand{\LS}{\ \ \ \ \ \ \ \ \ \ }
\newcommand{\ls}{\ \ \ \ \ }
\newcommand{\wt}{\widetilde}
\newcommand{\wh}{\widehat}
\newcommand{\ve}{\varepsilon}
\newcommand{\ol}{\overline}

\newcommand{\bsubeq}{\begin{subequations}}
\newcommand{\esubeq}{\end{subequations}}

\renewcommand{\d}{{\rm d}}

\newcommand{\R}{{\cal R}}

\newcommand{\nn}{\nonumber}

\newcommand{\e}{{\rm e}}

\newcommand{\hG}{{\wh{\Gamma}}}

\newcounter{Enumerate}

\newcommand{\para}{\parallel}

\begin{document}

\allowdisplaybreaks{

\setcounter{page}{0}

\begin{titlepage}

{\normalsize
\begin{flushright}
KEK-TH-903, OU-HET 453\\
hep-th/0307193\\
July 2003 
\end{flushright}
}

\vspace{1mm}

\begin{center}
{\huge Spectrum of Eleven-dimensional Supergravity

\vspace{3mm}

on a PP-wave Background}

\vspace{5mm}

\bigskip
{\large Tetsuji {\sc Kimura}$^{1,2}$ \ \ and \ \ Kentaroh {\sc Yoshida}$^1$}

\vspace{4mm}

{\sl
1: Theory Division, Institute of Particle and Nuclear Studies,\\
High Energy Research Organization (KEK)\\
Tsukuba, Ibaraki 305-0801, Japan

\vspace{5mm}

2: Department of Physics,
Graduate School of Science, Osaka University\\
Toyonaka, Osaka 560-0043, Japan

\vspace{5mm}

\tt tetsuji@post.kek.jp \ \ {\rm and} \ \ \tt kyoshida@post.kek.jp}

\end{center}

\vspace{4mm}


\begin{abstract}
We calculate the spectrum of the linearized supergravity 
around the maximally supersymmetric pp-wave background in eleven
dimensions. 
The resulting spectrum agrees with that of 
zero-mode Hamiltonian of a supermembrane theory on the 
pp-wave background.  We also discuss 
the connection with 
the Kaluza-Klein zero modes of $AdS_4 \times S^7$ background.  
\end{abstract}

\vspace{2cm}

\end{titlepage}


\newpage

\section{Introduction}

Recently, superstring and supermembrane on pp-wave 
backgrounds 
are focused upon. 
It was shown in \cite{M} that the type IIB superstring
theory on the maximally supersymmetric pp-wave background \cite{IIB:pp}
is exactly solvable, 
and the spectrum of type IIB string theory was
compared with the result calculated from the viewpoint of
the linearized supergravity around the pp-wave background \cite{MT}. 

In addition, a matrix model on maximally supersymmetric
pp-wave background was proposed in \cite{BMN}. 
By using the matrix regularization \cite{dWHN}, 
this matrix model is closely related to the
supermembrane theory on the pp-wave background \cite{DSR,SY1}. 
(For the works related to the matrix model and supermembrane theory 
on pp-waves, see \cite{alg} or \cite{SY2}.) 
In the paper \cite{NSY} the spectrum of
the zero-mode Hamiltonian in the supermembrane theory 
on the pp-wave background 
was obtained by following the method of 
\cite{dWHN} and solving the system quantum mechanically. 
Motivated by the work \cite{NSY}, 
we calculate the spectrum of the linearized
supergravity around the maximally supersymmetric pp-wave solution in
eleven dimensions \cite{KG}.  
In the flat case, 
it is known that 
the spectrum of the zero-mode Hamiltonian 
corresponds to that of the
linearized supergravity around Minkowski spacetime, 
and hence such a correspondence should be
shown in the case of pp-wave.  
In this scenario, we will see that the spectrum is given by 
$({\bf 1} + {\bf 28} + {\bf 35}) \times 2 \, ( \, = {\bf 128})$ 
for graviton and three-form gauge field (bosons), 
and $({\bf 8} + {\bf 56}) \times 2 \, ( \, = {\bf 128})$ 
for gravitino (fermions), 
and then
that the spectrum obtained from the linearized supergravity agrees with
that of the zero-mode Hamiltonian. 
Notably, the resulting spectrum is identical
with the Kaluza-Klein zero-modes in the eleven-dimensional supergravity
on seven-sphere $S^7$ \cite{Duff} as noted in \cite{NSY}.  

This paper is organized as follows: 
In section 2 
we prepare classical field equations in eleven-dimensional supergravity 
on the maximally supersymmetric pp-wave background. 
The Hamiltonian 
of Klein-Gordon type field equations for the fluctuation fields is also
discussed. 
In section 3, we derive linearized field equations for fluctuation
fields around the pp-wave background
and calculate the spectrum of graviton, three-form
gauge field and gravitino. The resulting spectrum agrees with that of
the zero-mode Hamiltonian in the supermembrane theory on pp-wave
background obtained in \cite{NSY}. 
We find that the spectrum corresponds to 
the Kaluza-Klein zero-modes in the supergravity on $S^7$ \cite{Duff}. 
Section 4 is devoted to conclusion and discussion.  
In Appendix A, our notation and convention are summarized.


\section{Setup}

We are interested in the spectrum of the linearized supergravity around 
the maximally supersymmetric pp-wave background (often called 
Kowalski-Glikman (KG) background \cite{KG}). In order to carry out our
calculation, we shall introduce the classical equations of motion in 
eleven-dimensional supergravity. For the later convenience, we also 
discuss the energy eigenvalue of the Hamiltonian of the fluctuation
fields. 

Eleven-dimensional supergravity consists of 
metric $g_{MN}$ (or vielbein $e_M{}^{\dot{A}}$), 
three-form gauge field $C_{MNP}$ and gravitino $\Psi_M$ 
as dynamical fields.
Equations of motion for these fields are\footnote{
The convention and field contents on supergravity action is denoted in
Appendix \ref{convention}.} 
\bsubeq \label{eom-g} 
\begin{align}
0 \ &= \half g_{MN} \R - \R_{MN} 
- \frac{1}{96} g_{MN} F_{PQRS} F^{PQRS}
+ \frac{1}{12} F_{MPQR} F_{N}{}^{PQR} 
\; , \\ 
0 \ &= \ 
\hG{}^{MNP} D_N \Psi_P 
- \frac{1}{96} \wt{\Gamma}{}^{MNPQRS} \Psi_N F_{PQRS} \; , \\
0 \ &= \ 
\nabla^Q \big\{ e \, F_{QMNP} \big\}
+ \frac{18}{(144)^2} \, g_{MZ} \, g_{NK} \, g_{PL} \, \ve^{ZKLQRSUVWXY} 
\, F_{QRSU} \, F_{VWXY}
\; ,
\end{align}
\esubeq
where we wrote down the equation of motion in terms of $g_{MN}$.  
We also neglected the terms derived from torsion, 
quadratic and higher order terms with respect to gravitino.  
Such higher order terms do not contribute to 
field equations of fluctuation modes in the first-order approximation. 


Now let us discuss the Hamiltonian and its energy eigenvalue. 
We need to calculate and solve field equations for fluctuation modes 
around the KG background (for the KG background, see Appendix
\ref{pp}) in the next section. 
Then we will encounter Klein-Gordon type equations of motion and 
have to evaluate its energy spectrum.

We shall consider a Klein-Gordon type equation of motion 
for a field $\phi (x)$:
\begin{align}
\big( \Box - \alpha \, \mu \, i \del_- \big) \phi (x^+, x^-, x^I) \ &= \ 0 
\; , \label{eom}
\end{align}
where $\alpha$ is an arbitrary constant and $x^+$ is an evolution
parameter. The d'Alembertian $\Box$ on the KG background is given by 
\begin{align}
\Box \ &= \ - \frac{1}{\sqrt{-g}} \del_M \big( \sqrt{-g} g^{MN} \del_N \big)
\ = \ 
2 \del_+ \del_- + G_{++} (\del_-)^2 - (\del_K)^2 
\; .
\end{align}
The above Klein-Gordon type field equation 
will appear later as equations of motion of fluctuation modes. 
Fourier transformed expression of $\phi (x)$
\begin{align*}
\phi (x^+, x^-, x^I) \ &= \ 
\int \! \frac{\d p_- \d^9 p_I}{\sqrt{(2 \pi)^{10}}}
\, \e^{i (p_- x^- + p_I x^I)} \, \wt{\phi} (x^+, p_-, p_I)
\end{align*}
leads to the following expression:
\begin{align}
0 \ &= \ 2 p_- \, i \del_+ - \wt{G}{}_{++} (p_-)^2 + (p_I)^2 
+ \alpha \, \mu \, p_-
\; ,
\end{align}
where $\wt{G}{}_{++}$ is defined as 
\begin{align}
\wt{G}{}_{++} \ &\equiv \ 
\sum_{\wt{I}=1}^3 \Big( \frac{\mu}{3} \Big)^2 (\del_{p_{\wt{I}}})^2
+ \sum_{I'=4}^9 \Big( \frac{\mu}{6} \Big)^2 (\del_{p_{I'}})^2
\; .
\end{align}
By rewriting the above equation and $H = i \del_+$, 
we can obtain the explicit expression of Hamiltonian: 
\begin{align}
H \ &= \ \frac{1}{-2 p_-} \big\{
(p_I)^2 - \wt{G}{}_{++} (p_-)^2 + \alpha \, \mu \, p_- \big\}
\; .
\label{hamiltonian}
\end{align}
The energy spectrum of this Hamiltonian can be derived by using the
standard technique of harmonic oscillators. 
Now we define ``creation/annihilation'' operators  
\bsubeq
\begin{align}
a^{\wt{I}} \ &\equiv \ 
\frac{1}{\sqrt{2 \wt{m}}} 
\big\{ p_{\wt{I}} + \wt{m} \del_{p_{\wt{I}}} \big\}
\; , &
\ol{a}{}^{\wt{I}} \ &\equiv \ 
\frac{1}{\sqrt{2 \wt{m}}} 
\big\{ p_{\wt{I}} - \wt{m} \del_{p_{\wt{I}}} \big\}
\; , &
\wt{m} \ &\equiv \ - \frac{1}{3} \mu \, p_- 
\; , \\
a^{I'} \ &\equiv \ 
\frac{1}{\sqrt{2 m'}} 
\big\{ p_{I'} + m' \del_{p_{I'}} \big\}
\; , &
\ol{a}{}^{I'} \ &\equiv \ 
\frac{1}{\sqrt{2 m'}} 
\big\{ p_{I'} - m' \del_{p_{I'}} \big\}
\; , &
m' \ &\equiv \ - \frac{1}{6} \mu \, p_- 
\; ,
\end{align}
\esubeq
whose commutation relations are represented by 
\begin{align*}
[ a^{\wt{I}} , \ol{a}{}^{\wt{J}} ] \ &= \ \delta^{\wt{I} \wt{J}}
\; , \ls
[ a^{I'} , \ol{a}{}^{J'} ] \ = \ \delta^{I'J'}
\; , \ls 
[ a^{\wt{I}} , \ol{a}{}^{J'} ] \ = \ 
[ a^{I'} , \ol{a}{}^{\wt{J}} ] \ = \ 0 
\; .
\end{align*}
Thus we express the Hamiltonian in terms of the above oscillators:
\begin{align}
H \ &= \ 
\frac{1}{3} \mu \sum_{\wt{I}} \ol{a}{}^{\wt{I}} a^{\wt{I}}
+ \frac{1}{6} \mu \sum_{I'} \ol{a}{}^{I'} a^{I'}
+ \frac{1}{2}\mu\left(2 - \alpha \right)
\; .
\end{align}
Note that the last term implies the zero-mode energy $E_0$ 
of the system, which is represented by
\begin{align}
E_0 \ &= \ \half \mu \, {\cal E}{}_0 (\phi) \; , 
\ls
{\cal E}{}_0 (\phi) \ = \ 2 - \alpha 
\; . \label{zero-energy}
\end{align}
In the next section, we will use ${\cal E}{}_0$ 
to evaluate the energy of the zero-modes of fluctuation fields.


\section{Spectrum on PP-wave Background}

In this section we discuss the spectrum of bosonic and fermionic
fields in eleven-dimensional supergravity on the pp-wave background.

Let us derive field equations for fluctuation fields. 
Fluctuation fields are expanded around classical fields as follows:
\begin{align}
g_{MN} \ &\to \ g_{MN} + h_{MN} 
\; , \ls
\Psi_M \ \to \ 0 + \psi_M 
\; , \ls
C_{MNP} \ \to \ C_{MNP} + {\cal C}{}_{MNP} 
\; , \label{f}
\end{align}
where $g_{MN}$ is the Kowalski-Glikman (KG) metric 
on the pp-wave background (\ref{KG-metric}) 
and $C_{MNP}$ satisfies $4 \del_{[+} C_{123]} = F_{+123} = \mu$.
Substituting (\ref{f}) into classical field equations (\ref{eom-g}),
we obtain linearized field equations for fluctuation fields around the pp-wave
background:
\bsubeq
\begin{align}
0 \ &= \ 
- \half g_{MN} \Big\{ 
h^{PQ} \, \R_{PQ} 
- \nabla^P \nabla^Q \, h_{PQ} 
+ \nabla^P \nabla_P \, h_{Q}{}^Q
\Big\}
\nn \\
\ & \ \ \ \ 
- \half \big\{
\nabla^P \nabla_M \, h_{NP}
+ \nabla^P \nabla_N \, h_{MP}
- \nabla_M \nabla_N \, h_{P}{}^P 
- \nabla^P \nabla_P \, h_{MN} 
\big\}
\nn \\
\ & \ \ \ \
- \frac{1}{24} g_{MN} \big\{
2 F^{PQRS} \, \del_{P} {\cal C}{}_{QRS}
- F_{PQRS} \, F_{U}{}^{QRS} \, h^{PU}
\big\}
\nn \\
\ & \ \ \ \ 
+ \frac{1}{3} F_M{}^{PQR} \, \del_{[N} {\cal C}{}_{PQR]} 
+ \frac{1}{3} F_N{}^{PQR} \, \del_{[M} {\cal C}{}_{PQR]} 
- \frac{1}{4} F_{MPQR} \, F_{NU}{}^{QR} \, h^{PU}
\; , 
\label{f-h} 
\end{align}
\begin{align}
0 \ &= \ 
\hG{}^{MNP} D_N \psi_P 
- \frac{1}{4} \mu \hG{}^{MN+123} \psi_N 
\nn \\ 
\ & \ \ \ \ 
- \frac{1}{4} \mu \Big\{
g^{M+} 
(\hG{}^{12} g^{3N} 
+ \hG{}^{23} g^{1N} 
+ \hG{}^{31} g^{2N})
- g^{M1}
(\hG{}^{23} g^{+N} 
+ \hG{}^{3+} g^{2N} 
+ \hG{}^{+2} g^{3N}) 
\nn \\
\ & \LS \ls
+ g^{M2} 
(\hG{}^{3+} g^{1N} 
+ \hG{}^{+1} g^{3N} 
+ \hG{}^{13} g^{+N})
- g^{M3}
(\hG{}^{+1} g^{2N} 
+ \hG{}^{12} g^{+N} 
+ \hG{}^{2+} g^{1N})
\Big\} \psi_N
\; , 
\label{f-psi} 
\end{align}
\begin{align}
0 \ &= \ 
4 g^{QR} \Big\{
\del_R \del_{[Q} {\cal C}{}_{MNP]}
- \Gamma^{S}_{RQ} \, \del_{[S} {\cal C}{}_{MNP]}
- \Gamma^{S}_{RM} \, \del_{[Q} {\cal C}{}_{SNP]}
- \Gamma^{S}_{RN} \, \del_{[Q} {\cal C}{}_{MSP]}
- \Gamma^{S}_{RP} \, \del_{[Q} {\cal C}{}_{MNS]}
\Big\}
\nn \\
\ & \ \ \ \ 
- \half g^{QR} \Big\{
F_{SMNP} 
(\nabla_R \, h_{Q}{}^S + \nabla_Q \, h_R{}^S - \nabla^S \, h_{RQ})
+ F_{QSNP} 
(\nabla_R \, h_{M}{}^S + \nabla_M \, h_R{}^S - \nabla^S \, h_{RM})
\nn \\
\ & \LS \ls \ \
+ F_{QMSP} 
(\nabla_R \, h_{N}{}^S + \nabla_N \, h_R{}^S - \nabla^S \, h_{RN})
+ F_{QMNS} 
(\nabla_R \, h_{P}{}^S + \nabla_P \, h_R{}^S - \nabla^S \, h_{RP})
\Big\}
\nn \\
\ & \ \ \ \ 
+ \frac{1}{144} g_{MZ} \, g_{NK} \, g_{PL} \, 
\ve^{ZKLQRSUVWXY} \, F_{QRSU} \, \del_V {\cal C}{}_{WXY}
\; . 
\label{f-c}
\end{align}
\esubeq
In the following subsection we will solve the above equations 
for the fluctuations by using some
gauge-fixing conditions. 


\subsection{Physical Modes of Bosonic Fields} 

Here the bosonic spectrum is derived by using the light-cone gauge-fixing:
$h_{-M} = {\cal C}{}_{-MN} = 0$. All we have to do is to 
solve various physical and unphysical modes from fluctuation
field equations.

First, we find a traceless condition 
\begin{align}
0 \ &= \ h_{II}  \label{h---1.5}
\end{align}
from the $(--)$ component of (\ref{f-h}).
Substituting (\ref{h---1.5}) into the $(-I)$ component of (\ref{f-h})
leads to
\begin{align}
h_{I+} \ &= \ \frac{1}{\del_-} \del_J h_{IJ} \; ,
\label{h-non-dyn}
\end{align}
and so we see that $h_{I+}$ is non-dynamical.
In the same way, we can read off the following condition 
from the $(+-I)$ component of (\ref{f-c}):
$\del_J {\cal C}{}_{+IJ} = 0$. 
The $(-IJ)$ component of (\ref{f-c}) leads to
the expression for the field ${\cal C}{}_{+IJ}$:
\begin{align}
{\cal C}{}_{+IJ} \ = \ \frac{1}{\del_-} \del_K {\cal C}{}_{IJK} 
\; ,
\end{align}
and hence we see that ${\cal C}{}_{+IJ}$ is also non-dynamical.
Under the above conditions,
the trace of (\ref{f-h}) 
gives an expression of a non-dynamical field $h_{++}$:
\begin{align}
h_{++} \ &= \ 
\frac{1}{(\del_-)^2} \del_I \del_J h_{IJ}
- \frac{1}{3 \del_-} \mu \, {\cal C}{}_{123} 
\; . \label{h++non-dyn}
\end{align}
In contrast to the IIB supergravity case \cite{MT}, $h_{++}$ includes the term
proportional to $\mu$. The appearance of this term is characteristic of 
our case \footnote{The spectrum of type IIA string theory and linearized
supergravity is studied in \cite{KS}. In this case $h_{++}$ contains 
the additional term proportional to $\mu$. }. 

By the use of the light-cone gauge-fixing conditions and 
the above-mentioned conditions for the non-dynamical modes, 
we can reduce field equations for $h_{MN}$ and ${\cal C}_{MNP}$ as follows:
\bsubeq 
\begin{align}
\mbox{$(\wt{I} \wt{J})$ component of (\ref{f-h})} : &&
0 \ &= \ 
\Box \, h_{{\wt{I}}{\wt{J}}} 
+ \frac{2}{3} \mu \, \delta_{{\wt{I}}{\wt{J}}} \, \del_{-} {\cal C}
\; ,
\label{h-II2} \\
\mbox{$(\wt{I} J')$ component of (\ref{f-h})} : &&
0 \ &= \ 
\Box \, h_{{\wt{I}}{J'}} 
+ \mu \, \del_{-} {\cal C}{}_{\wt{I} J'} 
\; ,
\label{h-II'2} \\
\mbox{$(I' J')$ component of (\ref{f-h})} : &&
0 \ &= \ 
\Box \, h_{{I'}{J'}} 
- \frac{1}{3} \mu \, \delta_{{I'}{J'}} \del_{-} {\cal C}
\; .
\label{h-I'I'2} \\
\mbox{$(\wt{I} \wt{J} \wt{K})$ component of (\ref{f-c})} : &&
0 \ &= \ 
\Box \, {\cal C}
- 2 \mu \, \del_- h_{\wt{I} \wt{I}} 
\; , 
\label{c-III2} \\
\mbox{$(\wt{I} \wt{J} K')$ component of (\ref{f-c})} : &&
0 \ &= \ 
\Box \, {\cal C}{}_{\wt{I} J'} - \mu \, \del_- h_{\wt{I} J'} 
\; , 
\label{c-III'2} \\
\mbox{$(\wt{I} J' K')$ component of (\ref{f-c})} : &&
0 \ &= \ 
\Box \, {\cal C}{}_{{\wt{I}}{J'}{K'}}
\; ,
\label{c-II'I'2} \\
\mbox{$(I'J'K')$ component of (\ref{f-c})} : &&
0 \ &= \ 
\Box \, {\cal C}{}_{I' J' K'}
- \frac{1}{6} \mu \, 
\ve^{I' J'K' W'X'Y'} \del_- {\cal C}{}_{W'X'Y'}
\; ,
\label{c-I'I'I'2}
\end{align}
\esubeq
where $\ve^{I'J'K'W'X'Y'}$ is the $SO(6)$ Levi-Civita symbol whose
normalization is $\ve^{456789} = \ve_{456789} =1$.
Note that 
we wrote the above equations in terms of the following 
two quantities defined by 
\begin{align*}
{\cal C}{}_{\wt{I} J'} \ &\equiv \ 
\half \ve_{\wt{I} \wt{K} \wt{L}} {\cal C}{}_{\wt{K} \wt{L} J'}
\; , \ls
{\cal C} \ \equiv \ 2 {\cal C}{}_{123} \; ,
\end{align*}
where we introduced the $SO(3)$ Levi-Civita symbol $\ve_{\wt{I} \wt{J}
  \wt{K}}$ ($\ve_{123} = \ve^{123} = 1$).


\vspace{5mm}

Now let us solve the above reduced equations of motion for fluctuation
modes, and derive the zero-mode energy spectrum and degrees of freedom of
bosonic fields.

First we consider the field ${\cal C}{}_{\wt{I} J'K'}$. 
{}From the above equation (\ref{c-II'I'2}),
we find that this field does not couple to the other fields. 
So the zero-mode energy ${\cal E}{}_0 ({\cal C}{}_{\wt{I} J' K'})$ 
and degrees of freedom ${\cal D} ({\cal C}{}_{\wt{I} J'K'})$ are given by
\begin{align}
{\cal E}_0 ({\cal C}{}_{\wt{I} J' K'}) \ &= \ 2
\; , \ls
{\cal D} ({\cal C}{}_{\wt{I} J' K'}) \ = \ 45 \; .
\label{energy-dof-C-II'I'}
\end{align}

Next, we consider $SO(3) \times SO(6)$ tensor fields $h_{\wt{I} J'}$
and ${\cal C}{}_{\wt{I} J'}$ coupled to each other. 
In order to diagonalize these coupled fields, we define
two complex fields $H_{\wt{I} J'}$ and $\ol{H}{}_{\wt{I} J'}$ as 
\begin{align}
H_{\wt{I} J'} \ &= \ h_{\wt{I} J'} + i {\cal C}{}_{\wt{I} J'} 
\; , \ls
\ol{H}{}_{\wt{I} J'} \ = \ h_{\wt{I} J'} - i {\cal C}{}_{\wt{I} J'} 
\; .
\end{align}
By using these fields, (\ref{h-II'2}) and (\ref{c-III'2}) can be
rewritten as 
\begin{align}
0 \ &= \ \big( \Box - \mu \, i \del_- \big) H_{\wt{I} J'}
\; , \ls
0 \ = \ \big( \Box + \mu \, i \del_- \big) \ol{H}{}_{\wt{I} J'}
\; .
\end{align}
Thus the zero-mode energies and degrees of freedom 
of $H_{\wt{I} J'}$ and $\ol{H}{}_{\wt{I} J'}$ are given by
\begin{align}
{\cal E}{}_0 (H_{\wt{I} J'}) \ &= \ 1
\; , \ls
{\cal E}{}_0 (\ol{H}{}_{\wt{I} J'}) \ = \ 3
\; , \ls
{\cal D} (H_{\wt{I} J'}) \ = \ {\cal D} (\ol{H}{}_{\wt{I} J'}) 
\ = \ 18
\; . \label{energy-dof-H-II'}
\end{align}

Then we will solve the field equations (\ref{h-II2}), (\ref{h-I'I'2}) and
(\ref{c-III2}) concerning $h_{\wt{I} \wt{J}}$, $h_{I'J'}$ and
${\cal C}$. Since these fields are coupled to one another, 
we have to diagonalize these fields in order to solve the equations. 
Hence let us introduce the following fields: 
\bsubeq
\begin{align}
\ls &&
h_{\wt{I} \wt{J}}^{\perp}
\ &\equiv \ 
h_{\wt{I} \wt{J}} - \frac{1}{3} \delta_{\wt{I} \wt{J}} \, h_{\wt{K} \wt{K}}
\; , &
h_{I' J'}^{\perp}
\ &\equiv \ 
h_{I' J'} - \frac{1}{6} \delta_{I' J'} \, h_{K' K'}
\; , && \ls \\
&&
h \ &\equiv \ h_{\wt{I} \wt{I}} + i {\cal C} 
\; , &
\ol{h} \ &\equiv \ h_{\wt{I} \wt{I}} - i {\cal C} \; .
\end{align}
\esubeq
Note that 
$h_{\wt{I} \wt{J}}^{\perp}$ and $h_{I' J'}^{\perp}$ are transverse modes
and two complex scalar fields $h$ and $\ol{h}$ are trace modes. 
In this re-definition we find $\Box h_{\wt{I} \wt{J}}^{\perp} = 0$, and
so its energy and 
degrees of freedom are given by 
\begin{align}
{\cal E}{}_0 (h_{\wt{I} \wt{J}}^{\perp}) \ &= \ 2 \; , \ls
{\cal D} (h_{\wt{I} \wt{J}}^{\perp}) \ = \ 5 
\; . \label{energy-dof-h-perp}
\end{align}
Since we also find $\Box h_{I' J'}^{\perp} = 0$,
we obtain the energy and degrees of freedom of zero-mode 
$h_{I' J'}^{\perp}$:
\begin{align}
{\cal E}{}_0 (h_{I' J'}^{\perp}) \ &= \ 2 \; , \ls
{\cal D} (h_{I' J'}^{\perp}) \ = \ 20
\; . \label{energy-dof-h'-perp}
\end{align}
Similarly the field equations for $h$ and $\ol{h}$ are described by
\begin{align}
\big( \Box - 2 \mu \, i \del_- \big) h \ &= \ 0 
\; , \ls 
\big( \Box + 2 \mu \, i \del_- \big) \ol{h} \ = \ 0 
\; .
\end{align}
Thus the energies and degrees of freedom of them are
\begin{align}
{\cal E}{}_0 (h) \ &= \ 0
\; , \ls
{\cal E}{}_0 (\ol{h}) \ = \ 4
\; , \ls
{\cal D} (h) \ = \ 
{\cal D} (\ol{h}) \ = \ 1 
\; . \label{energy-dof-h-hbar}
\end{align}

Finally we consider (\ref{c-I'I'I'2}) by decomposing ${\cal
  C}{}_{I'J'K'}$ into self-dual part and anti-self-dual part as follows:
${\cal C}{}_{I'J'K'} \equiv 
{\cal C}{}_{I'J'K'}^{\oplus} + {\cal C}{}_{I'J'K'}^{\ominus}$,
where 
${\cal C}{}_{I'J'K'}^{\oplus}$ is a self-dual part and 
${\cal C}{}_{I'J'K'}^{\ominus}$ is an anti-self-dual part. 
These are defined by, respectively, 
\begin{align}
{\cal C}{}_{I'J'K'}^{\oplus} \ &= \ 
\frac{i}{3!} \ve^{I'J'K'W'X'Y'} {\cal C}{}_{W'X'Y'}^{\oplus}
\; , \ls
{\cal C}{}_{I'J'K'}^{\ominus} \ = \ 
- \frac{i}{3!} \ve^{I'J'K'W'X'Y'} {\cal C}{}_{W'X'Y'}^{\ominus}
\; .
\end{align}
Due to this decomposition,
the field equations of them are expressed as
\begin{align}
\big( \Box + \mu \, i \del_- \big) {\cal C}{}_{I'J'K'}^{\oplus}
\ &= \ 0 
\; , \ls
\big( \Box - \mu \, i \del_- \big) {\cal C}{}_{I'J'K'}^{\ominus}
\ = \ 0 
\; ,
\end{align}
and hence we find the energies and degrees of freedom of 
${\cal C}{}_{I'J'K'}^{\oplus}$ and ${\cal C}{}_{I'J'K'}^{\ominus}$:
\begin{align}
{\cal E}{}_0 ({\cal C}{}_{I'J'K'}^{\oplus}) 
\ &= \ 3
\; , \ls
{\cal E}{}_0 ({\cal C}{}_{I'J'K'}^{\ominus})
\ = \ 1
\; , \ls
{\cal D} ({\cal C}{}_{I'J'K'}^{\oplus}) \ = \ 
{\cal D} ({\cal C}{}_{I'J'K'}^{\ominus}) \ = \ 
10
\; . \label{energy-dof-C-I'I'I'}
\end{align}

\vspace{5mm}

Now we have fully solved the field equations for bosonic fluctuations
and have derived the spectrum of graviton and three-form gauge field. 
The resulting spectrum is splitting with a certain energy difference 
in contrast to the flat case. The resulting spectrum is completely 
identical with that of zero-mode Hamiltonian 
in the supermembrane theory on the KG background 
when we take account of the difference between ${\cal E}{_0}$ and $E_0$.
We conclude this subsection
by summarizing the spectrum of bosonic fields in Table
\ref{boson}:

\begin{table}[htbp]
\begin{center}
\begin{tabular}{c|@{\ls}l@{\ls}l@{\ls}l@{\ls}|c} \hline
energy ${\cal E}{}_0$ 
& \multicolumn{3}{c|}{bosonic fields (${\cal D}$)} & degrees of freedom
\\ \hline \hline
$4$ & $\ol{h} (1)$ & & & $1$ \\
$3$ & $\ol{H}{}_{\wt{I} J'} (18)$ & ${\cal C}{}_{I'J'K'}^{\oplus}
  (10)$ & & $28$ \\
$2$ & ${\cal C}{}_{\wt{I} J'K'} (45)$ & $h_{\wt{I}
    \wt{J}}^{\perp} (5)$ & $h_{I'J'}^{\perp} (20)$ & $70$ \\
$1$ & $H_{\wt{I} J'} (18)$ & ${\cal C}{}_{I'J'K'}^{\ominus}
  (10)$ & & $28$ \\
$0$ & $h (1)$ & & & 1 \\ \hline 
\end{tabular}
\caption{\sl Bosonic zero-modes in eleven-dimensional supergravity on
  pp-wave background.}
\label{boson}
\end{center}
\end{table}


\subsection{Physical Modes of Fermionic Fields}

In this subsection let us solve the field equations of 
the fluctuations of gravitino by imposing the light-cone gauge-fixing:
$\psi_- = 0$. We will study  physical modes of gravitino and 
obtain its spectrum. 

First, we consider the $M=-$ component of (\ref{f-psi}), which is rewritten as
\begin{align}
\hG^N D_- \psi_N \ &= \ 
\frac{1}{9} \hG_{-} \hG_{N} J^N \; ,
\end{align}
where $J^N$ is defined as
\bsubeq
\begin{align}
\hG^{MNP} D_{N} \psi_P \ &\equiv \ J^M \; , \\
J^+ \ &= \ - J_- \ = \ 0 \; , \\
J^- \ &= \ - \frac{1}{4} \mu \, \hG^{+-123I'} \psi_{I'}
- \half \mu \, \ve_{\wt{I} \wt{J} \wt{K}} \, \hG_{\wt{I} \wt{J}} 
\psi_{\wt{K}} \; , \\
J^{\wt{I}} \ &= \ J_{\wt{I}} \ = \ 
\frac{1}{4} \mu \, \hG^{+123} 
\Big( \delta_{\wt{I} \wt{J}} - \hG_{\wt{I}} \hG_{\wt{J}} \Big) 
\psi_{\wt{J}} \; , \\
J^{I'} \ &= \ J_{I'} \ = \ 
- \frac{1}{4} \mu \, \hG^{+123} \Big( \delta_{I'J'}
- \hG_{I'} \hG_{J'} \Big) \psi_{J'} \; .
\end{align}
\esubeq
Then we can simplify as $\hG^N D_- \psi_N = \del_- ( \hG^N \psi_N)$,
and find that $\del_- ( \hG^N \psi_N)$ vanishes. Thus we obtain
\begin{align}
\hG^N \psi_N \ &= \ 0
\; .
\end{align}

Next, we consider the $M=+$ component of (\ref{f-psi}), 
which can be rewritten as
\begin{align}
0 \ &= \ g^{P+} \hG{}^N D_N \psi_P 
- g^{PN} \hG{}^+ D_N \psi_P
+ \half \big( 
\hG{}^+ \hG{}^N  - \hG{}^N \hG{}^+ 
\big) \hG{}^P D_N \psi_P
\; . \label{psi-+2}
\end{align}
Then the first and third term is deleted by the light-cone gauge-fixing, 
and we can reduce (\ref{psi-+2}) to
$0 = \hG{}^+ ( - \del_- \psi_+ + \del_I \psi_I)$.
Thus $\psi_+$ can be expressed as 
\begin{align}
\psi_+ \ &= \ \frac{1}{\del_-} \del_I \psi_I 
\; ,
\label{non-dyn-psi+} 
\end{align}
and we see that $\psi_+$ is a non-dynamical field. 


\vspace{5mm}

Here we shall reduce 
the $M=\wt{I}$ component of (\ref{f-psi}) to
\begin{align}
0 \ &= \ 
\hG{}^{\dot{+}} \Big( \del_+ + \half G_{++} \del_- \Big)
\psi_{\wt{I}}^{\oplus} 
+ \hG{}^{\dot{-}} \del_- \psi_{\wt{I}}^{\ominus}
+ \hG{}^{\dot{K}} \del_K 
(\psi_{\wt{I}}^{\oplus} + \psi_{\wt{I}}^{\ominus})
- \frac{1}{4} \mu \hG{}^{\dot{+}\dot{1}\dot{2}\dot{3}} 
\big( \delta_{\dot{\wt{I}} \dot{\wt{J}}} 
- \hG{}_{\dot{\wt{I}}} \hG{}_{\dot{\wt{J}}} \big) 
\psi_{\wt{J}}^{\oplus} 
\; , \label{psi-I4}
\end{align}
where 
we decomposed gravitino as $\psi_{\wt{I}} \equiv \psi_{\wt{I}}^{\oplus} +
\psi_{\wt{I}}^{\ominus}$. The $\psi_{\wt{I}}^{\oplus}$ and 
$\psi_{\wt{I}}^{\ominus}$ are defined as 
\begin{align}
\psi_{\wt{I}}^{\oplus} \ &\equiv \ - \half \hG{}^{\dot{-}}
\hG{}^{\dot{+}} \psi_{\wt{I}} \; , \ls 
\psi_{\wt{I}}^{\ominus} \ \equiv \ - \half \hG{}^{\dot{+}}
\hG{}^{\dot{-}} \psi_{\wt{I}} \; , \label{left-right-fermion}
\end{align}
which satisfy the projection conditions: 
$\hG{}^-\psi_{\wt{I}}^{\oplus} = \hG{}^+ \psi_{\wt{I}}^{\ominus} = 0$.
When we act $\hG{}^{\dot{+}}$ on (\ref{psi-I4}) from the left, 
$\psi_{\wt{I}}^{\ominus}$ can be expressed in terms
$\psi_{\wt{I}}^{\oplus}$ as follows:
\begin{align}
\psi_{\wt{I}}^{\ominus} \ &= \ 
\frac{1}{2 \del_-} \hG{}^{\dot{+}} \hG{}^{\dot{K}}
\del_K \psi_{\wt{I}}^{\oplus} 
\; . \label{non-dyn--fermion}
\end{align}
Thus $\psi_{\wt{I}}^{\ominus}$ is not independent of $\psi_{\wt{I}}^{\oplus}$.
Similarly, when we act $\hG{}^{\dot{-}}$ on (\ref{psi-I4}) from the left 
and utilize (\ref{non-dyn--fermion}), 
we obtain the following equation:
\begin{align}
0 \ &= \ 
\Box \, \psi_{\wt{I}}^{\oplus} 
- \half \mu \hG{}^{\dot{1}\dot{2}\dot{3}} 
\big( \delta_{\dot{\wt{I}} \dot{\wt{J}}} 
- \hG{}_{\dot{\wt{I}}} \hG{}_{\dot{\wt{J}}} \big) 
\del_- \psi_{\wt{J}}^{\oplus}
\; . \label{pre-eom-psiI}
\end{align}
In order to solve this equation, we shall introduce the following
fields: 
\begin{align}
\psi_{\wt{I}}^{\oplus \perp} \ &\equiv \ 
\Big( \delta_{\dot{\wt{I}} \dot{\wt{J}}} 
- \frac{1}{3} \hG{}_{\dot{\wt{I}}} \hG{}_{\dot{\wt{J}}}
\Big) \psi_{\wt{J}}^{\oplus} 
\; , \ls
\psi_1^{\oplus \para} \ \equiv \ 
\hG{}^{\dot{\wt{I}}} \psi_{\dot{\wt{I}}}^{\oplus} \ = \ 
\hG{}^{\dot{\wt{I}}} \psi_{\wt{I}}^{\oplus}
\; , \label{perp-para-psiI}
\end{align}
and decompose $\psi_{\wt{I}}^{\oplus}$ in to the $\hG$-transverse mode and 
$\hG$-parallel mode. 
Acting $\hG{}^{\dot{\wt{I}}}$ on (\ref{pre-eom-psiI}) from the
left and contracting the index ${\dot{\wt{I}}}$,
we get
\begin{align}
0 \ &= \ 
\Box \, \psi_1^{\oplus \para} - \mu \hG{}^{\dot{1}\dot{2}\dot{3}} 
\del_- \psi_1^{\oplus \para} 
\; . \label{para-eom-psiI}
\end{align}
On the other hand, 
when we act $(\delta_{\dot{\wt{K}} \dot{\wt{I}}} - \frac{1}{3}
\hG{}_{\dot{\wt{K}}} \hG{}_{\dot{\wt{I}}})$ on
(\ref{pre-eom-psiI}), we find  
\begin{align}
0 \ &= \ 
\Box \, \psi_{\wt{K}}^{\oplus \perp}
- \half \mu \hG{}^{\dot{1}\dot{2}\dot{3}} \del_- \psi_{\wt{K}}^{\oplus \perp} 
\; . \label{perp-eom-psiI}
\end{align}
Moreover, in order to solve (\ref{para-eom-psiI}) and 
(\ref{perp-eom-psiI}), 
we decompose $\psi_{\wt{I}}^{\oplus \perp}$ 
and $\psi_1^{\oplus \para}$ according to the chirality in terms of 
$i\hG{}^{\dot{1}\dot{2}\dot{3}} $ as follows: 
\bsubeq \label{lr-perp-para-fermion}
\begin{align}
\psi_{\wt{I} {\rm R}}^{\oplus \perp} 
\ &\equiv \ 
\frac{1 + i \hG{}^{\dot{1}\dot{2}\dot{3}}}{2} \psi_{\wt{I}}^{\oplus \perp}
\; , \ls
\psi_{\wt{I} {\rm L}}^{\oplus \perp} 
\ \equiv \ 
\frac{1 - i \hG{}^{\dot{1}\dot{2}\dot{3}}}{2} \psi_{\wt{I}}^{\oplus \perp}
\; , \\
\psi_{1 {\rm R}}^{\oplus \para} 
\ &\equiv \ 
\frac{1 + i \hG{}^{\dot{1}\dot{2}\dot{3}}}{2} \psi_1^{\oplus \para}
\; , \ls
\psi_{1 {\rm L}}^{\oplus \para} 
\ \equiv \ 
\frac{1 - i \hG{}^{\dot{1}\dot{2}\dot{3}}}{2} \psi_1^{\oplus \para}
\; .
\end{align}
\esubeq
These variables satisfy the following chirality conditions:
\bsubeq
\begin{align}
i \hG{}^{\dot{1}\dot{2}\dot{3}} \psi_{\wt{I} {\rm R}}^{\oplus \perp}
\ &= \ 
+ \psi_{\wt{I} {\rm R}}^{\oplus \perp} 
\; , \ls
i \hG{}^{\dot{1}\dot{2}\dot{3}} \psi_{\wt{I} {\rm L}}^{\oplus \perp}
\ = \ 
- \psi_{\wt{I} {\rm L}}^{\oplus \perp} 
\; , \\
i \hG{}^{\dot{1}\dot{2}\dot{3}} \psi_{1 {\rm R}}^{\oplus \para}
\ &= \ 
+ \psi_{1 {\rm R}}^{\oplus \para} 
\; , \ls
i \hG{}^{\dot{1}\dot{2}\dot{3}} \psi_{1 {\rm L}}^{\oplus \para}
\ = \ 
- \psi_{1 {\rm L}}^{\oplus \para} 
\; .
\end{align}
\esubeq
Multiplying $\half (1 \pm i \hG{}^{\dot{1}\dot{2}\dot{3}})$ to
(\ref{para-eom-psiI}) on the left, we get 
\begin{align}
0 \ &= \ 
\Big( \Box + \mu \, i \del_- \Big) 
\psi_{1 {\rm R}}^{\oplus \para}
\; , \ls
0 \ = \ 
\Big( \Box - \mu \, i \del_- \Big)
\psi_{1 {\rm L}}^{\oplus \para}
\; . \label{eom-psiI-1}
\end{align}
Similarly,
when we multiply $\half (1 \pm i \hG{}^{\dot{1}\dot{2}\dot{3}})$ to
(\ref{para-eom-psiI}) on the left, we obtain
\begin{align}
0 \ &= \ 
\Big( \Box + \half \mu \, i \del_- \Big)
\psi_{\wt{I} {\rm R}}^{\oplus \perp}
\; , \ls
0 \ = \ 
\Big( \Box - \half \mu \, i \del_- \Big)
\psi_{\wt{I} {\rm L}}^{\oplus \perp}
\; . \label{eom-psiI-2}
\end{align}
{}From these equations, 
we can read off the zero-mode energies and degrees of freedom of
$\psi_{\wt{I} {\rm R}}^{\oplus \perp}$ and $\psi_{\wt{I} {\rm
    L}}^{\oplus \perp}$:
\begin{align}
{\cal E}{}_0 (\psi_{\wt{I} {\rm R}}^{\oplus \perp}) 
\ &= \ \frac{5}{2} 
\; , \ls
{\cal E}{}_0 (\psi_{\wt{I} {\rm L}}^{\oplus \perp}) 
\ = \ \frac{3}{2}
\; , \ls
{\cal D} (\psi_{\wt{I} {\rm R}}^{\oplus \perp}) 
\ = \ {\cal D} (\psi_{\wt{I} {\rm L}}^{\oplus \perp}) 
\ = \ 
8 \times (3-1) \ = \ 16
\; . \label{energy-dof-psiI-perp}
\end{align}
We will discuss these quantities of $\psi_{1 {\rm R}}^{\oplus \para}$
and $\psi_{1 {\rm L}}^{\oplus \para}$ later. 


\vspace{5mm}

Then let us investigate the $M=I'$ component of (\ref{f-psi}):
\begin{align}
0 \ &= \ 
\Big\{ \hG{}^{\dot{+}} \Big( \del_+ + \half G_{++} \del_- \Big)
+ \hG{}^{\dot{-}} \del_- + \hG{}^{\dot{K}} \del_K \Big\}
\psi_{I'}
+ \frac{1}{4} \mu \hG{}^{\dot{+}\dot{1}\dot{2}\dot{3}} 
\Big( \delta_{\dot{I'}\dot{J'}} -
\hG{}_{\dot{I'}} \hG{}_{\dot{J'}} \Big) \psi_{J'}
\; . \label{psi-I'2}
\end{align}
In the same way as the case of $\psi_{\wt{I}}$, 
we decompose $\psi_{I'}$ into the $\hG$-parallel mode 
and $\hG$-transverse mode, and obtain 
\bsubeq
\begin{align}
0 \ &= \ 
\Big( \Box - \frac{5}{2} \mu \, i \del_- \Big)
\psi_{2 {\rm R}}^{\oplus \para}
\; , \ls
0 \ = \ 
\Big( \Box + \frac{5}{2} \mu \, i \del_- \Big)
\psi_{2 {\rm L}}^{\oplus \para}
\; , \label{eom-psiI'-1} \\
0 \ &= \ 
\Big( \Box - \half \mu \, i \del_- \Big) 
\psi_{I' {\rm R}}^{\oplus \perp}
\; , \ls
0 \ = \ 
\Big( \Box + \half \mu \, i \del_- \Big)
\psi_{I' {\rm L}}^{\oplus \perp}
\; , \label{eom-psiI'-2}
\end{align}
\esubeq
where the $\hG$-transverse mode and $\hG$-parallel mode 
are defined as
\bsubeq \label{left-right-fermion-I'}
\begin{align}
\psi_{I'}^{\oplus} \ &= \ 
- \half \hG{}^{\dot{-}} \hG{}^{\dot{+}} \psi_{I'} 
\; , \\
\psi_{I' {\rm R}}^{\oplus \perp} 
\ &= \ 
\frac{1 + i \hG{}^{\dot{1}\dot{2}\dot{3}}}{2} \psi_{I'}^{\oplus \perp}
\; , \ls
\psi_{I' {\rm L}}^{\oplus \perp} 
\ = \ 
\frac{1 - i \hG{}^{\dot{1}\dot{2}\dot{3}}}{2} \psi_{I'}^{\oplus \perp}
\; , \\
\psi_{2 {\rm R}}^{\oplus \para} 
\ &= \ 
\frac{1 + i \hG{}^{\dot{1}\dot{2}\dot{3}}}{2} \psi_{2}^{\oplus \para}
\; , \ls
\psi_{2 {\rm L}}^{\oplus \para} 
\ = \ 
\frac{1 - i \hG{}^{\dot{1}\dot{2}\dot{3}}}{2} \psi_{2}^{\oplus \para}
\; .
\end{align}
\esubeq
{}From (\ref{eom-psiI'-2}),
we find that the zero-mode energies and degrees of freedom of 
are given by 
$\psi_{\wt{I} {\rm R}}^{\oplus \perp}$ and $\psi_{\wt{I} {\rm
    L}}^{\oplus \perp}$:
\begin{align}
{\cal E}{}_0 (\psi_{I' {\rm R}}^{\oplus \perp}) 
\ &= \ \frac{3}{2} 
\; , \ls
{\cal E}{}_0 (\psi_{I' {\rm L}}^{\oplus \perp}) 
\ = \ \frac{5}{2}
\; , \ls
{\cal D} (\psi_{I' {\rm R}}^{\oplus \perp}) 
\ = \ {\cal D} (\psi_{I' {\rm L}}^{\oplus \perp}) 
\ = \ 
8 \times (6-1) \ = \ 40
\; . \label{energy-dof-psiI'-perp}
\end{align}


\vspace{5mm}

Now we have finished the study of the $\hG$-transverse part. 
The remaining task is to analyze the $\hG$-parallel mode. 
In order to investigate the $\hG$-parallel mode, 
we perform a linear combination of (\ref{eom-psiI-1}) and
(\ref{eom-psiI'-1}), and define new $\hG$-parallel modes as 
\begin{align}
\psi_{\rm R}^{\oplus \para} \ &\equiv \ 
\frac{2}{5} \psi_{1 {\rm R}}^{\oplus \para} 
- \psi_{2 {\rm R}}^{\oplus \para}
\; , \ls
\psi_{\rm L}^{\oplus \para} \ \equiv \ 
\frac{2}{5} \psi_{1 {\rm L}}^{\oplus \para} 
- \psi_{2 {\rm L}}^{\oplus \para}
\; . 
\end{align}
Then, by the use of (\ref{non-dyn-psi+}), 
 we can easily see that the re-defined fermions satisfy the equations:
\begin{align}
0 \ &= \ 
\Big( \Box - \frac{3}{2} \mu \, i \del_- \Big) \psi_{\rm R}^{\oplus
  \para}
\; , \ls
0 \ = \ 
\Big( \Box + \frac{3}{2} \mu \, i \del_- \Big) \psi_{\rm L}^{\oplus
  \para}
\; . \label{eom-para}
\end{align}
Thus the zero-mode energies and degrees of freedom of them are 
represented by 
\begin{align}
{\cal E}{}_0 (\psi_{\rm R}^{\oplus \para})
\ &= \ 
\half
\; , \ls
{\cal E}{}_0 (\psi_{\rm L}^{\oplus \para})
\ = \
\frac{7}{2} 
\; , \ls
{\cal D} (\psi_{\rm R}^{\oplus \para})
\ = \ 
{\cal D} (\psi_{\rm L}^{\oplus \para})
\ = \ 8 
\; . \label{energy-dof-psi-para}
\end{align}

Now we have fully solved the field equations for fermionic
fluctuations, and have derived the spectrum of gravitino in the case of
pp-wave. As a result, we have found that the spectrum is splitting 
with a certain energy difference in the same manner with the spectrum of 
bosons.  
Summarizing (\ref{energy-dof-psiI-perp}),
(\ref{energy-dof-psiI'-perp}) and (\ref{energy-dof-psi-para}),
we obtain the spectrum of gravitino as in Table
\ref{fermion}:

\begin{table}[htbp]
\begin{center}
\begin{tabular}{c|@{\ls}l@{\LS}l@{\ls}|c} \hline
energy ${\cal E}{}_0$ & \multicolumn{2}{c|}{fermionic fields (${\cal D}$)} 
& degrees of freedom
\\ \hline \hline
$\frac{7}{2}$ & $\psi_{\rm L}^{\oplus \para} (8)$ & & $8$ \\
$\frac{5}{2}$ & $\psi_{\wt{I} {\rm R}}^{\oplus \perp} (16)$ & 
$\psi_{I' {\rm L}}^{\oplus \perp} (40)$ & $56$ \\
$\frac{3}{2}$ & $\psi_{\wt{I} {\rm L}}^{\oplus \perp} (16)$ & 
$\psi_{I' {\rm R}}^{\oplus \perp} (40)$ & $56$ \\
$\frac{1}{2}$ & $\psi_{\rm R}^{\oplus \para} (8)$ & & $8$ \\ \hline
\end{tabular}
\caption{\sl Fermionic zero-modes in eleven-dimensional supergravity on
  pp-wave background.}
\label{fermion}
\end{center}
\end{table}
This spectrum completely agrees with that of the zero-mode Hamiltonian 
in the supermembrane theory on the KG background \cite{NSY}.  

In conclusion, we have shown that the spectrum of the linearized
supergravity around the KG background is completely identical with that 
of the zero-mode Hamiltonian in the supermembrane theory on this
background \cite{NSY}. 
It should be remarked that the spectrum obtained here is 
also identical with the Kaluza-Klein zero-modes in the supergravity on
$S^7$. The $AdS_4\times S^7$ background is related to the KG background 
\cite{BFHP2} via the Penrose limit \cite{P}. Hence the spectrum on the
KG background would have a connection with that on the $AdS_4\times
S^7$. In fact, the relationship of the spectra between the $AdS_4\times
S^7$ and KG background is discussed from the viewpoint of superalgebra 
\cite{FGP}.


\section{Conclusion and Discussions}

We have calculated the spectrum of the linearized supergravity around
the maximally supersymmetric pp-wave background. 
As a result, we have found that the spectrum is given by 
$({\bf 1} + {\bf 28} + {\bf 35}) \times 2 \, (\, ={\bf 128})$ 
for graviton and three-form gauge field (bosons), and 
$({\bf 8} + {\bf 56}) \times 2 \, (\, ={\bf 128})$ 
for gravitino (fermions), 
and then that the spectrum obtained from the linearized supergravity 
agrees with that of the zero-mode Hamiltonian. 
Notably, the resulting spectrum is identical with the Kaluza-Klein 
zero-modes in the eleven-dimensional supergravity 
on seven-sphere $S^7$. 

Our considerations in this paper have clarified the spectrum of the
supergravity multiplet splitting with a certain energy difference.  
As an application of our results, 
it would be possible to 
calculate propagators and energy-momentum tensors of graviton, 
gauge field and gravitino according to our classification of matter
contents by following the method of \cite{prop}. 
In contrast to the flat case, 
the polarization tensor is non-trivially decomposed. 
Hence we need a concrete classification of physical fields 
in order to calculate the propagators and energy-momentum tensors. 
If the calculation of them has been completed, 
we can evaluate the scattering amplitudes 
in the linearized supergravity around the pp-wave background. 
It is one of the most important problems to
calculate the scattering amplitude in the matrix model and to compare the
result with the amplitudes obtained from the supergravity side. 
Thus, we believe that our work will make an important contribution 
to study in this direction. 
We will continue to study in this direction and calculate the
propagators and energy-momentum tensors \cite{future}.

\section*{Acknowledgements}

We would like to thank Makoto Sakaguchi for valuable comments.
The work of T.K. is supported in part by JSPS Research Fellowships for
Young Scientists.


\section*{Appendix}

\begin{appendix}

\section{Convention} \label{convention}

In this appendix we summarize some conventions.

\subsection{Convention of gamma matrices}

We use the convention that $M,N,P, \cdots$ refer to
eleven-dimensional world indices and $\dot{A},\dot{B},\dot{C}, \cdots$ 
refer to eleven-dimensional tangent space indices.
The eleven-dimensional gamma matrices satisfy the following relations
\bsubeq
\begin{align}
\{ \hG{}^{\dot{A}} , \hG{}^{\dot{B}} \} \ &= \ 
2 \eta^{\dot{A}\dot{B}} 
\; , \\
(\hG{}^{\dot{A}})^{\dagger} \ &= \ \Gamma_{\dot{A}} \ = \ 
- \hG{}^{\dot{0}} \hG{}^{\dot{A}} 
(\hG{}^{\dot{0}})^{-1} 
\; , \\
\hG{}_{M_1 M_2
  \cdots M_n} \ &\equiv \ \hG{}_{[M_1} \hG{}_{M_2} \cdots
  \hG{}_{M_n]} \ = \ \frac{1}{n!} {\rm sgn} (\sigma)
  \hG{}_{M_{\sigma(1)}} \hG{}_{M_{\sigma(2)}} \cdots
  \hG{}_{M_{\sigma(n)}} 
\; ,
\end{align}
\esubeq
where $\eta_{\dot{A}\dot{B}}$ is the tangent space metric.
Irreducible spinors in eleven-dimensional tangent space 
are anti-commuting variables and satisfy the Majorana condition:
\begin{align}
\ol{\Psi} \ &= \ \Psi^T C \; , 
\end{align}
where the charge conjugation matrix $C$ is antisymmetric and defined
by
\bsubeq
\begin{align}
C \hG{}^{\dot{A}} C^{-1} \ &= \ - (\hG{}^{\dot{A}})^T \; , \\
C \hG{}^{\dot{A}_1 \cdots \dot{A}_{2n}} C^{-1} \ &= \ 
- ( \hG{}^{\dot{A}_1 \cdots \dot{A}_{2n}} )^T \; , \\
C \hG{}^{\dot{A}_1 \cdots \dot{A}_{2n+1}} C^{-1} \ &= \ 
( \hG{}^{\dot{A}_1 \cdots \dot{A}_{2n+1}} )^T \; .
\end{align}
\esubeq


\subsection{Eleven-dimensional supergravity}

Eleven-dimensional supergravity contains graviton, three-form gauge
field and gravitino. The degrees of freedom is {\bf 128} (bosons) + 
{\bf 128} (fermions). 
These fields are described by 
\bsubeq
\begin{align}
e_M{}^{\dot{A}} \ &: \ \ \mbox{vielbein} \; , \ls 
E_{\dot{A}}{}^M \ : \ \ \mbox{inverse vielbein} \; , \\
\Psi_M \ &: \ \ \mbox{gravitino (Majorana spinor)} \; , \\
C_{MNP} \ &: \ \ \mbox{completely antisymmetric tensor} \; .
\end{align}
\esubeq
The Lagrangian of eleven-dimensional supergravity (without torsion) 
is given by   
\begin{align}
{\cal L} \ &= \ 
e \R 
- \half e \, \ol{\Psi}{}_M \, \hG{}^{MNP} \, D_N \Psi_P 
- \frac{1}{48} e \, F_{MNPQ} \, F^{MNPQ} \nn \\
\ & \ \ \ \
+ \frac{1}{192} e \, \ol{\Psi}{}_M \, \wt{\Gamma}{}^{MNPQRS} \, \Psi_N
\, F_{PQRS} 
+ \frac{1}{(144)^2} \ve^{MNPQRSUVWXY} \, F_{MNPQ} \, F_{RSUV} \, C_{WXY}
\; , \label{SUGRA}
\end{align}
where the covariant derivative for local Lorentz transformation is
defined as
\begin{align*}
D_N \Psi_P \ &= \ 
\del_N \Psi_P - \frac{1}{4} \omega_N{}^{\dot{A}\dot{B}} 
\hG{}_{\dot{A}\dot{B}} \Psi_P
\; . 
\end{align*}
We also introduced the following expressions:
\begin{align*}
g_{MN} \ = \ e_M{}^{\dot{A}} \, e_N{}^{\dot{B}} \,
\eta_{\dot{A}\dot{B}} 
\; , \ls 
\eta_{\dot{A}\dot{B}} 
\ &= \ 
E_{\dot{A}}{}^M \, E_{\dot{B}}{}^N \, g_{MN} \; , \ls
e \ = \ \det e_M{}^{\dot{A}} \; , \\
\ol{\Psi} \ &= \ i \Psi^{\dagger} \hG{}^0 \; ,
\end{align*}
and $\wt{\Gamma}{}^{NPQR}{}_M$ and $\wt{\Gamma}{}^{MNPQRS}$ are 
written as 
\begin{align*}
\wt{\Gamma}{}^{NPQR}{}_M \ &= \ 
\hG{}^{NPQR}{}_M 
- 8 \delta_M{}^{[N} \hG{}^{PQR]} \; , \ls
\wt{\Gamma}{}^{MNPQRS} \ = \ 
\hG{}^{MNPQRS} + 12 g^{M[P} \hG{}^{QR} g^{S]N} \; .
\end{align*}
The eleven-dimensional Levi-Civita symbol, 
$\ve^{MNPQRSTUVWX}$ is 
normalized as 
\begin{align*}
\ve^{012\cdots10} \ &= \ 1 \; .
\end{align*} 
The Lagrangian written above is invariant under the 
supersymmetry transformations (up to torsion):
\bsubeq
\begin{align}
\delta e_{M}{}^{\dot{A}} \ &= \ 
- \half \ol{\ve} \hG{}^{\dot{A}} \Psi_M \; , \\
\delta \Psi_M \ &= \ 
2 D_M \ve + \frac{1}{144} F_{NPQR} (\wt{\Gamma}{}^{NPQR}{}_M \ve) \; , \\
\delta C_{MNP} \ &= \ 
- \frac{3}{2} \ol{\ve} \hG{}_{[MN} \Psi_{P]} \; . 
\end{align}
\esubeq


\subsection{Kowalski-Glikman background} \label{pp}

Here we will summarize several properties of 
the maximally supersymmetric pp-wave background. 
This solution was found by Kowalski-Glikman \cite{KG} and 
often called the KG solution. This is the unique pp-wave type 
solution preserving maximal supersymmetries. 
The metric of this background is given by 
\begin{align}
\d s^2 \ &= \ - 2 \d x^+ \d x^- + G_{++} (\d x^+)^2 + \sum_{I=1}^9 (
\d x^{I})^2 \; , \label{KG-metric} \\
G_{++} \ &= \ - \Big[ \Big( \frac{\mu}{3} \Big)^2 \sum_{\wt{I}=1}^3
  (x^{\wt{I}})^2 + \Big( \frac{\mu}{6} \Big)^2 \sum_{I'=4}^9
  (x^{I'})^2 \Big] \; , \nonumber 
\end{align}
which is equipped with the constant flux 
\[
 \mu \ = \ F_{+123} \ \neq \ 0 \; . 
\]
Here the light-cone coordinates are defined as 
\begin{align*}
x^{\pm} \ &= \ \frac{1}{\sqrt{2}} \big( x^0 \pm x^{10} \big) \; .
\end{align*}

Affine connection $\Gamma^{P}_{MN}$, Riemann curvature
tensor $R^R{}_{PMN}$ and spin connection $\omega_M{}^{\dot{A} \dot{B}}$ 
are defined by 
\begin{align*}
\Gamma^{P}_{MN} \ &= \ \half g^{PR} \big( 
\del_M g_{NR} + \del_N g_{MR} - \del_R g_{MN} \big) \; , \\
R^R{}_{PMN} \ &= \ 
\del_M \Gamma^R_{PN} - \del_N \Gamma^R_{PM} 
+ \Gamma^R_{QM} \Gamma^Q_{PN} - \Gamma^R_{QN} \Gamma^Q_{PM} \; , \\
\omega_M{}^{\dot{A}\dot{B}} \ &= \ 
- \eta^{\dot{B}\dot{C}} E_{\dot{C}}{}^N \big\{ \del_M \, e_N{}^{\dot{A}} -
\Gamma^P_{NM} \, e_P{}^{\dot{A}} \big\} \; .
\end{align*}
In our consideration the contribution from torsion is not included, i.e.,
affine connection is symmetric under lower indices: $\Gamma^P_{MN} =
\Gamma^P_{NM}$. 
For the KG metric, the above quantities are written as 
\bsubeq
\begin{align}
\Gamma^{\wt{I}}_{++} \ &= \ \Big( \frac{\mu}{3} \Big)^2 x^{\wt{I}} 
\ = \ - \half \del^{\wt{I}} G_{++}
\; , &
\Gamma^{I'}_{++} \ &= \ \Big( \frac{\mu}{6} \Big)^2 x^{I'} 
\ = \ - \half \del^{I'} G_{++}
\; , \\
\Gamma^{-}_{+\wt{I}} \ &= \ \Gamma^{-}_{\wt{I}+} \ = \ 
\Big( \frac{\mu}{3} \Big)^2 x^{\wt{I}} 
\ = \ \Gamma^{\wt{I}}_{++}
\; , &
\Gamma^{-}_{+I'} \ &= \ \Gamma^{-}_{I'+} \ = \ 
\Big( \frac{\mu}{6} \Big)^2 x^{I'} 
\ = \ \Gamma^{I'}_{++}
\; , \\
R^{\wt{J}}{}_{+\wt{I}+} \ &= \ 
\delta_{\wt{I} \wt{J}} \Big(\frac{\mu}{3} \Big)^2 \;, &
R^{J'}{}_{+I'+} \ &= \ 
\delta_{I' J'} \Big(\frac{\mu}{6} \Big)^2 \;, \\
\omega_+{}^{\dot{\wt{I}}\dot{-}} 
\ &= \ \Big( \frac{\mu}{3} \Big)^2 x^{\wt{I}} \; , &
\omega_+{}^{\dot{I'}\dot{-}} 
\ &= \ \Big( \frac{\mu}{6} \Big)^2 x^{I'} \; .
\end{align}
\esubeq 
It should be noted that the scalar curvature vanishes and 
the Ricci tensor is constant and proportional to $\mu^2$. 
These are given by 
\begin{eqnarray}
\R_{++}  &=&  \R^{--}  \;=\;  \half \mu^2 \;, \quad \R  \;=\; 0 \; . 
\end{eqnarray}

\end{appendix}



}
\end{document}